\documentclass[runningheads]{llncs}
\usepackage{algorithm,algorithmic}
\usepackage{amsmath}
\usepackage{graphicx}
\usepackage{enumitem}
\usepackage{float}
\usepackage{gensymb}
\usepackage[font=normalsize,skip=6pt]{caption}
\usepackage{color}   
\usepackage{soul}
\usepackage[normalem]{ulem}
\usepackage[bookmarks,bookmarksopen,bookmarksdepth=2]{hyperref}
\usepackage[pscoord]{eso-pic}

\usepackage[utf8]{inputenc}
\usepackage[T1]{fontenc}

\newcommand{\placetextbox}[3]{
  \setbox0=\hbox{#3}
  \AddToShipoutPictureFG*{
    \put(\LenToUnit{#1\paperwidth},\LenToUnit{#2\paperheight}){\vtop{{\null}\makebox[0pt][c]{#3}}}%
  }%
}%

\begin{document}

\placetextbox{0.2}{0.96}{\color{red}Author's version}%
\placetextbox{0.7}{0.96}{\color{red}Constructive Side-Channel Analysis and}%
\placetextbox{0.7}{0.94}{\color{red}Secure Design - COSADE 2021}%
\placetextbox{0.7}{0.9}{\color{red}DOI:10.1007/978-3-030-89915-8\_1}

\date{}

\title{\Large \bf SideLine: How Delay-Lines (May) Leak Secrets from your SoC}
\titlerunning{SideLine}

\author{
Joseph Gravellier\inst{1} \and
Jean-Max Dutertre\inst{2} \and
Yannick Teglia\inst{1} \and
Philippe Loubet Moundi\inst{1}
}

\authorrunning{J. Gravellier et al.}

\institute{Thales, La Ciotat, France\\
\email{\{name.surname\}@thalesgroup.com}\\\and
Mines Saint-Etienne, CEA-Tech, Centre CMP. Gardanne, France\\
\email{dutertre@emse.fr}}

\maketitle 

\begin{abstract}
To meet the ever-growing need for performance in silicon devices, SoC providers have been increasingly relying on software-hardware cooperation. By controlling hardware resources such as power or clock management from the software, developers earn the possibility to build more flexible and power efficient applications.
Despite the benefits, these hardware components are now exposed to software code and can potentially be misused as open-doors to new kind of attacks.
In this work, we introduce \textit{SideLine}, a novel side-channel vector based on delay-line components widely implemented in high-end SoCs. 
We demonstrate that these entities can be used to perform remote power side-channel attacks 
and we 
detail several 
attack scenarios in which an adversary process located in one processor core aims at eavesdropping the activity of a victim process located in another core. For each scenario, we demonstrate the adversary ability to fully recover the secret key of an AES algorithm running in the victim core. Even more detrimental, we show that these attacks are still practicable 
when a rich operating system is used.
\end{abstract}

\section{Introduction}
   
The need for direct physical access to a target to perform a hardware attack was recently proved obsolete. Software-exposed hardware mechanisms implemented to improve SoC performance 
or power consumption 
were shown to be susceptible to remote hijacking by attackers seeking to perform fault injection or Side-Channel Attacks (SCAs).\par
Since 2014, and the \textit{Rowhammer} vulnerability's disclosure \cite{Kim2014}, the remote attack threat has become 
prevalent in hardware security researches. As a matter of fact, the influx of connected devices associated with the multiplication of cloud services offers a new playing field for attackers. Moreover, despite the appearance of trusted entities (ARM TrustZone, Intel SGX) that testify a growing need for SoC security, the hardware threat remains underestimated.\par  
Between 2014 and today, \textit{Rowhammer} capability evolved from random bit flips generation to privilege escalation on remote devices \cite{Gruss2015,Kurmus2017,Weissman2019}. Meanwhile, the \textit{CLKSCREW} exploit demonstrated that power and clock glitch attacks can be launched from within an ARM SoC using software programmable voltage-frequency regulators \cite{Tang2017}. Recently, this attack was improved \cite{Qiu2019} and deployed on Intel SGX devices \cite{Kenjar2019,Murdock2020}. From a side-channel point of view, two novel families of remote attacks have been introduced. On the one hand, micro-architectural timing attacks with \textit{Meltdown-Spectre} \cite{Lipp2018,Kocher2019}, \textit{Foreshadow} (SGX) \cite{VanBulck2018b} and more recently \textit{MDS} exploits \cite{Schaik2019,Canella2019}. These attacks leverage speculative and out-of-order execution in modern processors to steal secret data from victim processes. On the other hand, remote power SCAs have been introduced through several works on FPGA devices. Through the implementation of sensors inside a multi-user FPGA fabric, it was demonstrated that an adversary can eavesdrop the activity of the other users \cite{Schellenberg2018}. 
More recently, remote power SCAs have been extended to microcontroller devices using the ADCs they embed \cite{Gnad2019,OFlynn2019}
and to Intel devices using the RAPL interface \cite{Lipp2021Platypus}.
This spreads further the threats posed by remote SCAs from FPGA fabrics to general purpose microcontrollers as those found in usual connected devices.\par
In this paper we introduce \textit{SideLine}, a novel side-channel vector based on the intentional misuse of hardware resources available in high-end SoC devices. \textit{SideLine} leverages delay-lines components embedded in SoCs that use external memory; it neither requires embedded reconfigurable logic (FPGA) nor analog circuitry (ADC). Two delay-line blocks namely \textit{delay-locked-loop} and programmable \textit{delay-block} are hijacked to perform voltage measurements and maliciously used to conduct power SCAs on application processors (AP) and microcontrollers units (MCU). \textit{SideLine} makes it possible for an attacker to perform software-induced hardware attacks without direct physical access to the target. Our contributions are listed below:


\begin{enumerate}

    
    \item[$\bullet$] We reveal that delay-line-based components available in a broad range of SoCs that employ external memories can be turned into power consumption measurement units.
    
    
    
    \item[$\bullet$] We describe three attacker-victim (core-vs-core) delay-line-based SCA scenarios over two SoC devices: \textbf{AP-vs-AP} attack (on a Xilinx Zynq 7000 SoC), \textbf{AP-vs-MCU} attack and \textbf{MCU-vs-AP} attack (on a STMicroelectronics STM32MP1 SoC) where AP and MCU respectively denote the application processor and the microcontroller.

    \item[$\bullet$]  For each scenario a correlation power analysis attack is conducted against the publicly available OpenSSL AES encryption algorithm and the full secret key is successfully recovered. The attack feasibility is demonstrated on bare metal and Linux OS-based applications. 
    
\end{enumerate}    

\textbf{Responsible Disclosure:} We responsibly disclosed our findings to Xilinx on September 22th, 2020 and STMicroelectronics on November 2nd, 2020. Both acknowledged and agreed on the publication of these results. Moreover, this disclosure led to a close collaboration with these companies to find and build efficient countermeasures against \textit{SideLine} and similar attacks. Please keep in mind that \textit{SideLine} has been performed on these two processors for demonstration purposes but the concept is generic and any devices that embed delay-lines can be affected.\par
\textbf{Outline}: The remainder of this paper is organized as follows. In section \ref{sec:background}, we provide background information on power SCAs and describe the state-of-the-art. In section \ref{sec:SoC_Sensors}, we introduce delay-lines and their applications in SoC devices. Then, we present the tested products and the associated threat model in section \ref{sec:expsetup}. Sections \ref{sec:DLL} and \ref{sec:DelayBlockSCA} are dedicated to the deployment of the three attack scenarios. Finally, we discuss performance, limitations, countermeasures in section \ref{sec:Discussion} and conclude in section \ref{sec:conclusion}.

\begin{figure}[t]
\centering
\noindent
\includegraphics[width=11cm]{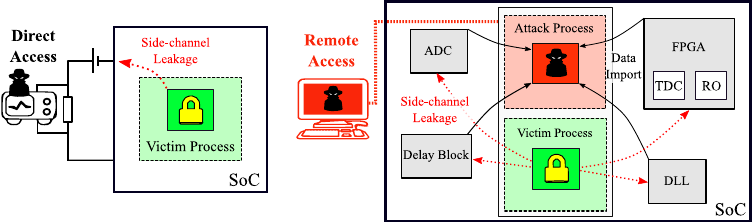}
\caption{On the left, local power SCA uses voltage probes to eavesdrop a leakage from a victim process. On the right, remote power SCA leverages the target's resources to monitor the victim process leakage without requiring physical access.}
\label{REMOTE}
\vspace{-3mm}
\end{figure}

\section{Background}\label{sec:background}

This section reminds the general side-channel background, the techniques recently introduced to monitor on-chip voltage fluctuations and the related works.


\subsection{Power Side-Channel Attacks}

A power SCA makes use of transistors switching activity leakage through power consumption variations to collect information about the processes running inside a device. Thanks to the correlation that exists between this leakage and the processed data, an attacker may try to launch an SCA to recover secret data or cryptographic keys from a target. Traditional power SCAs monitor the voltage variations induced by a device through a resistor attached to its power pads \cite{Kocher1999}. Simply by analysing the collected traces, an attacker can visually speculate on the different instructions executed by the target using a so-called Simple Power Analysis (SPA \cite{Kocher1999}) attack. Such SPA was proved effective to recover the private key used by asymmetric encryption algorithms like RSA or ECC \cite{Walter2004}.
Differential Power Analysis \cite{Kocher1999} and Correlation Power Analysis (CPA) \cite{Brier2010} use statistical tools to infer secret keys by correlating guessed leakage hypotheses with a set of experimental traces. 
\par Traditionally, power SCAs are carried out locally, in laboratories, using a voltage probe and an oscilloscope as depicted by the direct physical access attack path in Figure \ref{REMOTE}. These attacks target secure integrated circuits, such as smart cards or cryptographic accelerators embedded in SoCs. SCA countermeasures such as masking, jitter or shuffling \cite{Zhang2016,Zhou2005} are usually implemented in such secure devices. It encourages the use of high resolution and high sampling rate oscilloscopes on the attacker side to outperform the countermeasures.\par
Because traditional hardware attacks are assumed local 
and expensive, a large number of electronic devices are not prepared to withstand remote hardware attack scenarios. For this reason, even with limited performances, digital and analog integrated sensors may manage to jeopardize the security of devices ranging from IoT components to cloud servers (remote access in Figure \ref{REMOTE}). With the advent of these software-induced hardware attacks that do not require either direct physical access to the target or specific equipment, the alleged hardware attack limitations are called into question or even removed.

\subsection{On-Chip Voltage Sensing}


Two families of sensors enable malicious on-chip voltage sensing: either delay sensors built with digital logic gates which aim at measuring fluctuations in the power consumption through delay variations \cite{Zick2012,Zick2013}, or analog sensors using ADCs usually embedded in MCUs \cite{Gnad2019,OFlynn2019}. Until this work, digital sensors dedicated to SCAs have been exclusively implemented in FPGAs. Their available programmable logic makes it possible to design and tune such delay sensors in order to measure the power consumption of a device. We describe hereafter the principles of these delay sensors as their working principle is similar to the delay-line components we used.\par Delay-based voltage sensors leverage a side-effect of voltage fluctuations over digital logic behavior, which is the relationship between the time taken by a signal to propagate through a digital logic gate and the on-chip voltage level. An increase of the gate's power supply translates into a shortening of its propagation delay, and respectively a reduction of the voltage induces its increase \cite{Dutertre2012}. As a result, measuring the variations of the logic gates propagation delay provides an image of their voltage supply variations. Temperature and capacitive effects also play a significant part in its equation \cite{Dutertre2012}. Unlike voltage, the propagation delay can be directly measured using digital logic. Commonly used FPGA-based sensors are the Ring-Oscillator (RO \cite{Zick2012}) and the Time-to-Digital Converters (TDC \cite{Schellenberg2018}). \par 

\subsection{Related Works}

 In 2018, Schellenberg et al. demonstrated that FPGA-based sensors were precise enough to be used for SCAs on public and secret cryptographic algorithms \cite{Schellenberg2018}. To enable this attack, the adversary (a TDC-based delay sensor and its control logic for power supply measurement) and the victim (an AES hardware encryption block) needed to be located within the same FPGA. We define it as an \textbf{FPGA-to-FPGA} attack. The associated threat model targets multi-user FPGA cloud services that may appear over the next few years \cite{Chen2014}. 
 The same year, Zhao et al. disclosed that power SCAs can be conducted on heterogeneous platforms that include both an application processor and an FPGA fabric on the same silicon die. As a proof of concept, they were able to successfully retrieve the secret key of a custom RSA implementation running within a CPU core \cite{Zhao2018}. To do so, they carried out an SPA attack using RO-based voltage sensors implemented in the FPGA fabric.
 \par
Until 2019, remote power SCA remained bounded to FPGA devices or heterogeneous SoCs embedding an FPGA fabric as its flexibility allowed the implementation of powerful sensors. Two works went beyond the FPGA by proving that on-chip power SCAs can be carried out in microcontroller devices \cite{Gnad2019,OFlynn2019}. These attacks use ADCs as a straightforward way to measure on-chip power supply level. Thanks to a leakage of the chip power consumption into this analog block, the ADC can substitute the voltage probe role. Even with an extremely limited sampling rate, this noise sampling method was successful in retrieving the secret keys used by real world software and hardware AES cryptographic libraries.


\section{Delay-Lines in High-End SoC Devices}\label{sec:SoC_Sensors}

Delay-line-based sensors were previously used in FPGA devices as a way to monitor chip power consumption (TDC sensor). Despite offering great performance, these sensors were limited to configurable logic which is rarely integrated in SoC devices. In this section, we disclose that digital and analog delay-lines are widely implemented in SoC memory controllers. We present them and discuss their potential use as voltage sensors (delay sensors).

\begin{figure}[t]
    \centering
    \includegraphics[width=10cm]{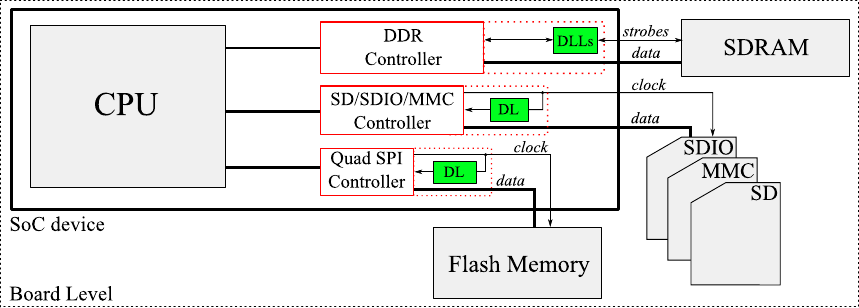}
    \caption{Typical SoC connectivity with external memories. Delay-lines are implemented to synchronize clock and data signals arrival in the memory controllers.}
    \label{MEMORY}
    \vspace{-3mm}
\end{figure}

\subsection{Memory Controller Basics}

Because high-end SoCs are designed to run operating systems (Linux, Android, etc.), they require a large amount of Non-Volatile Memory (NVM) to store the OS and Random-Access-Memory (RAM) to efficiently load it. 
Due to technological constraints, these SoCs do not embed a significant amount of RAM nor NVM memory but are rather interconnected with external memories  (memory cards, Flash memory, SDRAM memory, etc). Thus, depending on the form-factor, speed and memory size constraints, designers can choose between a wide range of external memory devices. A typical scenario of a SoC using external memories is depicted in Figure \ref{MEMORY}.\par
Several memory controllers are required to interface the SoC with its external memories. Each memory controller acts as a request arbiter, a transaction scheduler and as a physical interface to manage data flowing from the SoC to the memory, and vice-versa. In embedded systems, for cost and efficiency reasons, the memory controller is more likely to be directly integrated as a part of the SoC. At the edge of the memory controller, a physical controller (dotted lines in Figure \ref{MEMORY}) outputs and captures the signals that will flow between the SoC I/Os and the memory device I/Os (clock, data, configuration signals, etc.). The physical controller also ensures that these signals arrive on time regardless of the interconnection tracks length on the PCB, the voltage and the temperature variations. To better understand the extent of memory signal propagation timings, we draw a simple example of SoC/Synchronous Dynamic-RAM (SDRAM) association.
When a read operation is initiated by the SoC, the external SDRAM memory outputs the requested data edge-aligned with a clock signal (strobe) later dedicated to data sampling. Depending on the PCB tracks length, the clock signal 
is likely to shift 
ahead of the data signals, leading then to a sampling error. To mitigate this effect, the SoC physical controller implements delay-line-based components (delay-locked-loop \textit{DLL} and programmable delay-block \textit{DL} in Figure \ref{MEMORY}) to calibrate the phase alignment between the sampling clock and the data signals. This calibration can be manual and made once and for all after testing at manufacturing or performed at each chip power-up. It can also be adjusted dynamically to counterbalance any misalignment due to power supply or temperature fluctuations.\par
The relationship between the delay applied and the SoC voltage fluctuations drew our interest. In the following paragraphs, we present two different delay-line-based mechanisms that can be used to generate these delays for low and high-bandwidth external memory applications.


\subsection{Delay-blocks in Low-Bandwidth Memory Controllers}\label{sub:delayBlocks}

In relatively low-bandwidth external memories such as Flash memories, SD cards and multimedia cards, the impact of voltage and temperature fluctuations is considered not significant enough to jeopardize the communication integrity: dynamic calibration is not required. Delay-lines are nonetheless used to mitigate the impact of the PCB track length on the data and clock signals propagation timings (these delays are not predictable by SoC designers, they are set only at board design time). As track lengths are fixed, a static delay is sufficient to ensure good operation. For a read transaction, the delay-line is typically calibrated in order to add a phase shift of 90$^{\circ}$ to the clock signal. Thus, it ensures that data signals are in place when sampling occurs. The delay-line calibration is carried out through a series of training steps. These training steps modify the delay of the elements forming the chain and, for each configuration, verify if the external memory has been properly read. If the training is successful, the delay-line configuration is saved in a dedicated register and remains unchanged until the next test.\par

\begin{figure}[t]
\centering
\includegraphics[width=10cm]{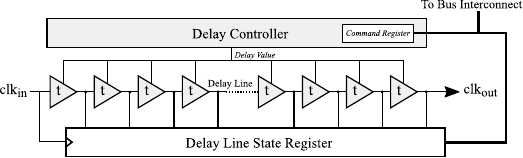}
\caption{An example of delay-block used in low-bandwidth memory controllers.}\label{DL}
\vspace{-3mm}
\end{figure}

Several SoC vendors provide user programmable delay-blocks as a way for developers to be able to use a wide range of memory chips or cards with different bus speeds. Unlike traditional static delay-lines, these delay-blocks come with both a complete calibration toolkit and a detailed documentation. Figure \ref{DL} illustrates the delay-block structure that was observed in one of the SoC we benchmarked. Its purpose is to delay the clock signal with respect to the data signals when a read operation is conducted. 
The block consists in a simple delay-line associated with a set of control and status registers. A \textit{Command Register} controls the delay $t$ of all the delay-line elements and thus the phase shift added to the $clk$ signal. To ensure that the phase shift obtained is conform to the applied command, a \textit{state register} captures the output of each element forming the delay-line every time a $clk_{in}$ rising edge event occurs. Then, a specific training is performed to verify whether the captured pattern matches the command or not.
\par Despite some missing parts, this structure is reminiscent of that of a TDC as the delay-line state is continuously captured and stored in an accessible register.  In section \ref{sec:DelayBlockSCA}, we demonstrate that this delay-block can be turned into a voltage sensor and hijacked to perform a power SCA.

%
%

\subsection{DLLs in High-Bandwidth Memory Controllers}\label{sub:delay-locked-loop}

Because of the continuous increasing in memory bus speeds, the available slack time for data sampling is gradually shrinking. Double data rate memories (DDR) such as SDRAM memory perform one data transfer per clock edge (both rising and falling) while reaching gigahertz frequencies \cite{Romo}. On these devices, the data sampling is very likely to get corrupted by 
temperature and voltage variations. This time, a static delay source is not suitable to ensure correct operations. To effectively cancel voltage and temperature noise side-effects, a dynamic way to adapt the clock delay has to be considered.

\begin{figure}[t]
\centering
\includegraphics[width=10cm]{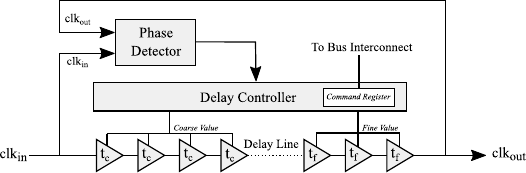}
\caption{An example of delay-locked-loop used in DDR memory controllers.}\label{DLL}
\vspace{-3mm}
\end{figure}

Delay Locked Loops (DLLs) are generally used in recent DDR memory controllers to dynamically track and control the phase shift applied between the sampling clock and the external memory (e.g. SDRAM) data signals \cite{ARM2003,Chung2007}. As illustrated in Figure \ref{DLL}, a DLL has two main blocks:
a delay-line, and a feedback circuit. The delay-line is calibrated to provide a phase shift to a \textit{clk} signal using both \textit{coarse} and \textit{fine} delay elements. However, the propagation delay jitter associated with on-chip voltage and temperature fluctuations is likely to skew the applied phase. This is why a DLL includes a feedback circuit to tune the delay-line in order to provide a dynamic control of the phase shift and thus, counterbalance voltage and temperature variations.
The feedback circuit comes with a phase detector that compares the phase shift between the clock signal at the input of the delay-line, $clk_{in}$, and its phase-shifted  clock output, $clk_{out}$. Then, according to the measured error, a delay controller applies a correction in order to "deskew" the result, that is, to get back to the initial delay. The applied correction modifies the delay of the elements forming the delay-line and can be either analog or digital-controlled depending on the delay-line type \cite{Abdulrazzaq2016a}. \par
A \textit{command register} stores the delay settings, it is memory-mapped and hence can be read from the SoC AP or MCU cores. The DLL operates autonomously, this means that through a simple access to this register, a process can retrieve the state of the DLL, which shall be correlated to on-chip voltage and temperature variations. As a result, tracking the \textit{command register} content shall provide an image of the SoC power consumption that may be used to carry out SCAs. Note that this measurement methodology (tracking the command of a feedback dynamically controlled system) differs from that described in Section \ref{sub:delayBlocks} for delay-blocks (sampling a clock signal propagating inside a fixed delay-line). 
If this unusual measurement medium provides enough resolution and sampling rate to eavesdrop power consumption of secure applications running on a processor, this could represent an important backdoor for computer security. This hypothetical vulnerability is strengthened by the fact that this attack only requires a read access to the command register, no configuration steps are required. This attack scenario is developed in section \ref{sec:DLL}.

\section{Experimental Setup}\label{sec:expsetup}


\subsection{Tested Devices}

Two devices from two different SoC providers have been studied in our experiments. 
The first target considered in this work is a Xilinx Zynq-7000 SoC \cite{XilinxInc2012} that comes with a dual-core Cortex-A9 application processor (AP). It is a typical multi-purpose SoC providing many additional resources: FPGA, I/O, ADCs, bus controllers, etc. It supports DDR2-DDR3, Flash and SD/MMC external memories and provides several DLL blocks to interface properly with DDR external memories. 
The experiments made on this target have been conducted without using an OS: we denote it as a \textbf{bare metal attack}. This configuration makes SCA easier as there are fewer interruptions (with respect to the case in which an OS is used) that may disturb the attack and victim processes and cause synchronization issues. The entire Zynq-based \textit{SideLine} attack code can be cloned from GitHub: \url{https://github.com/Remote-HWA/SideLine_Zynq}.
\par
The second target is a STMicroelectronics STM32MP157C-DK2 development board \cite{Microelectronics2019} that embeds a dual-core Cortex-A7 AP associated with a Cortex-M processor (MCU).  It also supports DDR2-DDR3, Flash and SD/MMC external memories and embeds several DLL blocks. Additionally, it provides user programmable delay-blocks (DLYB \cite{Microelectronics2019}) that can be employed for interfacing low bandwidth memory (e.g. an SD card). These programmable delay-blocks are the second case we studied. The experiments done on this SoC have been carried out with a Linux OS running on its AP (i.e. the Cortex-A7 processor). The results are those of a \textbf{Linux OS attack}. The entire STM32MP1-based \textit{SideLine} attack code can be cloned from GitHub: \url{https://github.com/Remote-HWA/SideLine_STM32MP1}.\par


\subsection{OpenSSL AES Architecture}

The OpenSSL library \cite{OpenSSL2002} provides several cryptographic algorithms used for securing channels over computer networks. In this work, we focus on the OpenSSL AES-128 (version 1.1.1) that implements a 32-bit tabulated version of the textbook AES encryption algorithm \cite{Rijmen1999}. This variant merges the {\tt Mixcolumn} and {\tt SubBytes} transformations into 4 pre-computed look-up tables known as T-tables (256 x 32-bit) as a way to optimize the computations on 32-bit processors. 


\subsection{Threat Model}\label{sub:ThreatModel}

In this work, we introduce three core-vs-core attack scenarios in order to assess the SCA capabilities of the delay-line-based sensors. For each scenario depicted in Figure \ref{THREAT_MODEL}, we first deploy a cryptographic application (in green) within a processor core. This application located either in the AP or in the MCU allows the end-user to launch AES encryptions/decryptions, with the plaintexts/ciphertexts that he provides. Secondly, we introduce a malicious user (in red) that has the privilege level necessary to access the delay-line blocks presented in Section \ref{sec:SoC_Sensors} and that uses them to retrieve the leakage induced by the AES application. \par
Although not used in this research work, Trusted Execution Environment (TEE) and TrustZone \cite{ARM2003} architecture stand as potential realistic targets for the delay-lines. TrustZone attacks from the normal-world to the secure-world have been widely covered in recent remote attack works \cite{Tang2017,Qiu2019,Bukasa2018,OFlynn2019}. However, from a side-channel point of view, the current TrustZone does not provide any countermeasures. Thus, the ability of an attacker to turn our feasibility attack into an end-to-end TrustZone attack is reasonably expected.\par
In the remainder of the paper, the three scenarios presented are referred to as: 
 


\begin{figure*}[t]
    \centering
    \includegraphics[width=12cm]{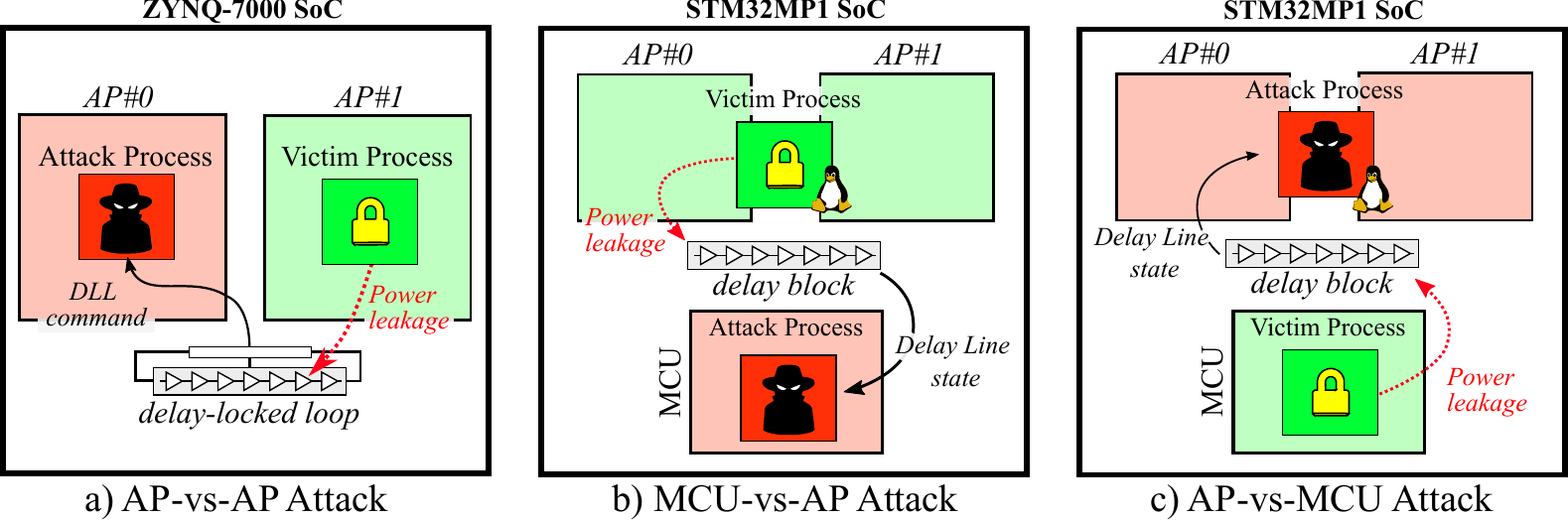}
    \caption{Basic principle of the three core-vs-core attack variants proposed in this work. It shows the leakage path from the victim process to the delay sensor and the sensor data flow retrieved by the attack process.}
    \label{THREAT_MODEL}
    \vspace{-5mm}
\end{figure*}

\begin{enumerate}
    \item A \textbf{DLL-based attack} (Figure \ref{THREAT_MODEL}.a), or AP-vs-AP attack, that demonstrates the ability of a DLL to serve as a power supply sensor suitable for a CPA attack against the AES algorithm. In this scenario, one core of the Zynq processor runs the AES victim application, while the second core executes the attack process (both victim and aggressor processes are C programs, in bare metal mode). The attacker code is in charge of collecting the leakage data of the AES. It does so by configuring the access to the DLL command register that makes it possible to sample its values during AES encryptions performed by the first core. The attacker core is also in charge of providing the plaintext to be ciphered by the victim process and to trigger both the encryption and readback of DLL states. This AP-vs-AP attack scenario is described in details in Section \ref{sec:DLL}.
    \item A first \textbf{Delay-Block-based attack} (Figure \ref{THREAT_MODEL}.b), or MCU-vs-AP attack, where the victim process is ran on the STM32MP1 AP (a C code AES running on top of a Linux OS) and the attack process is executed by the Cortex-M MCU (a C program, in bare metal mode). In this scenario the MCU is in charge of calibrating and using a delay-block to eavesdrop the activity of the AP. This MCU-vs-AP attack scenario is addressed in Section \ref{sec:DelayBlockSCA}.
    \item A second \textbf{Delay-Block-based attack} (Figure \ref{THREAT_MODEL}.c), or AP-vs-MCU attack, that matches a typical state-of-the-art industrial case where the cryptographic and security operations of a SoC embedding AP cores are delegated to a less complex MCU core. In this scenario the AP core (Cortex-A7) runs the attack process while the MCU core (Cortex-M) runs the AES victim process. This AP-vs-MCU attack scenario is reported in Section \ref{sec:DelayBlockSCA}.
\end{enumerate}

\section{DLL-based Power Side-Channel Attack}\label{sec:DLL}

This section presents a novel way to monitor on-chip voltage fluctuations and conduct power SCAs using the DLLs embedded in SoC memory controllers. 

\subsection{Validating DLL Effectiveness: \textit{Monitoring Temperature}}\label{sub:MonitorTemp}

\begin{figure}[t]
    \centering
    \includegraphics[width=10cm]{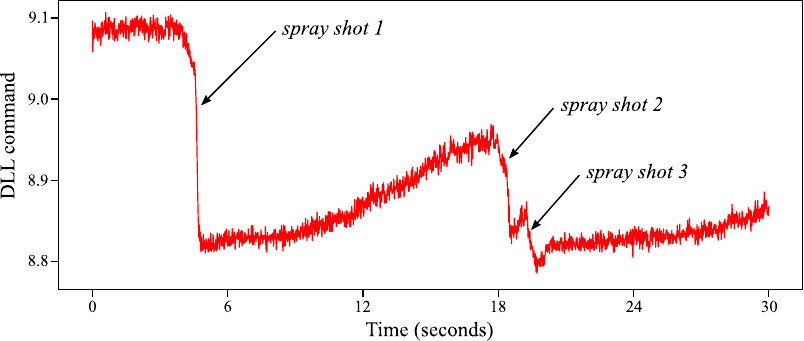}
    \caption{DLL response to sudden temperature drops induced by three successive exposition of the SoC to a cooling spray.}
    \label{temperature}
    \vspace{-5mm}
\end{figure}

As a proof of concept, a simple experiment was carried-out on the Zynq SoC to confirm that the DLL command is actually tracking the SoC package temperature variations. 
The test uses a C program designed to continuously read and store the DLL command register content into an acquisition array for a period of 30 seconds. Simultaneously, a cooling spray was used at specific moments to cool down the SoC package. To limit the acquisition size, each array index contains the average of 1,000 successive DLL readings. Figure \ref{temperature} reports the evolution of the measured DLL command 
(y-axis) as a function of time (x-axis). 
Each spray shot induces a temperature drop (translated into a DLL command drop in Figure \ref{temperature}) that progressively recovers until the next one.
This simple experiment confirms that a DLL is suitable to dynamically track the SoC temperature variations.
As the temperature decreases, the propagation speed of the $clk$ signal through the delay-line increases \cite{Dutertre2012}. Thus, the phase-shift between $clk_{in}$ and $clk_{out}$ progressively drifts. To counterbalance this effect, the DLL dynamically adapts its command in order to maintain a constant phase shift. Because package temperature evolves relatively slowly, the sampling frequency for this experiment was limited to 300 kHz. However, as this paper focuses on power side-channel, which itself depends on transient voltage drops measurements, a higher sampling rate needs to be achieved: it is the subject of the next subsection \ref{sub:DMA}.

\subsection{Improving Sampling Rate and Synchronisation using DMA}\label{sub:DMA}

As mentioned before, the DLL command value can be directly accessed through its memory address. Then, a loop associated with an array can be added to collect more samples. This CPU-based sampling method works in principle but has several drawbacks:\par
First, it requires a constant time between each acquisition. If this constant time is not achieved, the samples won't be correctly aligned. Consequently, statistical attacks will be less accurate as the averaging of several acquisitions will suffer from de-synchronisation. 
Achieving constant time is feasible in bare metal applications because they rarely suffer from interruptions. 
However, if the application runs over an OS, interrupts will dramatically affect the timing of acquisitions and make their averaging impossible.
The second limitation is related to the achievable sampling rate. Indeed, the delay induced by CPU memory access plus the storage of the acquired data into an array is not optimal. Using this method on the Zynq SoC, the sampling frequency was limited to 2.2 MHz.\par
To solve these issues, we choose to use Direct Memory Access (DMA) in order to improve the sampling rate as well as the synchronisation of our samples (as proposed in \cite{Gnad2019}). A DMA is a hardware module able to transfer data from a peripheral to another without processor intervention. For this reason, it is faster in transmitting data, but also not affected by OS interrupts. The source address (address from which the DMA should sample the data) is the register containing the DLL command. The destination address (destination of the DMA transfer) is the base address of an array whose size depends on the number of samples required. At the end of the DMA transfer, an interrupt flag is set and ends the sampling process. With DMA up and running, we improved the DLL sampling frequency from 2.2 MHz to 16 MHz.

%

\subsection{Bare Metal OpenSSL AES Attack Setup}\label{sub:AttackSetUpDLL}

According to the threat model we consider (see subsection \ref{sub:ThreatModel}), the attack process shall be able (1) to trigger the start of an AES encryption by the victim process, and (2) to control the gathering of the leakage from the AES through a DLL-based voltage sensor. Our test bench includes two processes (their pseudo codes are given in appendix \ref{CA9-0} and \ref{CA9-1}) executed by the two application cores of our target in bare metal mode: the attack process on AP\#0 and the victim process on AP\#1. \par

In addition to this attack setup, we used embedded hardware performance counters to precisely measure the duration of an AES encryption. On average, an encryption took 837 AP clock cycles or 1,25 $\mu$s  at a frequency of 667 MHz (both attack and victim programs were compiled with the optimization parameter set to -O2). The DMA transfer method we used provides a constant 62.5 ns sampling period (i.e. a 16 MHz sampling frequency). As a result, 21 samples of the DLL command are gathered per AES encryption.

\subsection{DLL-based SCA Attack on Zynq SoC}

\begin{figure}[t]
    \centering
    \includegraphics[width=\columnwidth]{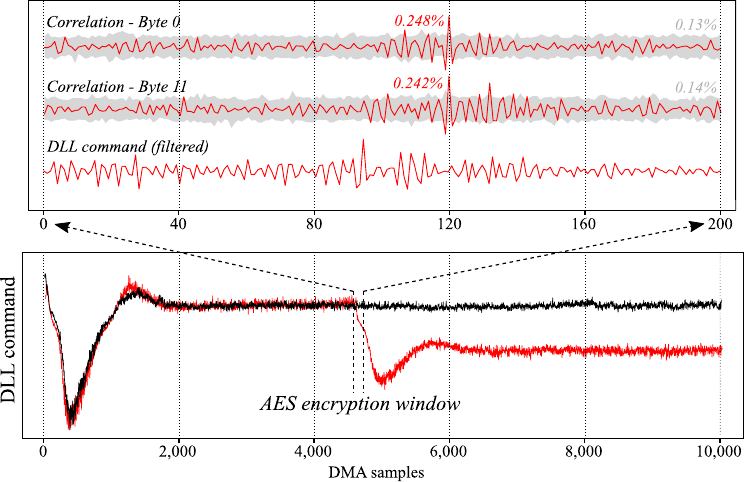}
    \caption{DLL-based attack results: the bottom part represents the impact of an AES encryption on the DLL command value. The top part zooms on the AES encryption windows and provides the temporal correlation rate for two key bytes.}
    \vspace{-5mm}
    \label{ZYNQ_attack}
\end{figure}

The bottom part of Figure \ref{ZYNQ_attack} illustrates the results of two experiments conducted to assess the AES encryption impact on the DLL command value and precisely detect its encryption time window.
The two traces depicted in black ($1^{st}$ case) and red ($2^{nd}$ case) represent the averaged DLL command value (y-axis) obtained for 1,000  acquisitions as a function of time (expressed in DMA samples). For the first experiment (in black), the victim program was kept idle during the entirety of the DMA sampling operations. The DLL command drop visible between sample 0 and 1,000 was induced by the extra power consumption linked to the DMA module activation. The DLL applied a strong correction to maintain a constant phase shift, that was finally relaxed as the power consumption returned to normal (sample 2,000 to the end of sampling). The second case (in red) reports an actual iteration of the attack and victim processes when an AES encryption is done. The red trace experienced the same DLL command undershoot due to DMA module activation (sample 0 to 1,000) but also a second undershoot corresponding to the AES encryption (starting at sample 4,500). It is finally restored to a steady value lower than the initial one (sample 6,000 to the end of sampling). The AES encryption window was deduced from the position of the second DLL command drop. Based on this information the CPA attack could be conducted on a smaller amount of samples. \par

We launched a total number of 20 million AES encryptions and acquired 200 DLL command samples per encryption. Samples and plaintexts extraction through UART took around 8 hours at 921,600 bauds. Then, an external computer was used to apply post-processing to the traces and conduct the CPA attack. The top part of Figure \ref{ZYNQ_attack} depicts a filtered and averaged trace of the DLL command (in red). High-pass filtering was used as a way to reduce the impact of low frequency variations (induced for instance by temperature fluctuations) on the acquired traces and thus to reduce the number of traces required for the attack. Then, we performed a plaintext-based CPA attack on the first round of the AES. As we mentioned earlier the OpenSSL AES uses T-tables to upgrade its performances on 32-bit processors. This allows us to leverage a 32-bit T-tables output prediction: $HW[T_{table}(key \oplus plaintext)]$. The obtained correlation results  versus the time are represented above the averaged trace in Figure \ref{ZYNQ_attack} (for two key bytes). The correct key hypotheses are depicted in red and emerge from the incorrect hypotheses (in grey) at sample 120. 
Based on 20 million encryptions, we achieved a full AES key recovery. 3 bytes were retrieved in the range 0-5M traces, 2 between 5-10M million, 5 between 10-15M an 4 between  15-20M. The key bytes number 7 and 9 never completely emerged from the incorrect candidates, but we assume that a simple brute force can be conducted to retrieve their values. The progressive correlation of the first 8 key bytes plus the failed byte \#9 are depicted in Figure \ref{Appendix:AP-vs-AP} in the appendix.

\subsection{Conclusion on DLL-based SCA}

In this section, we demonstrated that a DLL can be used to monitor on-chip temperature and power supply fluctuations. 
This unconventional voltage sensor was then used to conduct a power SCA on an OpenSSL AES algorithm implemented in the Zynq application processor and a full AES key recovery was achieved (with the help of brute force for the two remaining bytes). Performance, limitations and potential countermeasures regarding this attack are discussed in Section \ref{sec:Discussion}.

\section{Delay-Block-based Power Side-Channel Attack}\label{sec:DelayBlockSCA}

The DLL-based attack presented in Section \ref{sec:DLL} was associated with the use of DDR external memories such as SDRAM in AP-based SoC. 
This section discloses a second attack path that allows the hijacking of a programmable delay-block and its malicious use to perform core-vs-core power SCAs. These experiments are conducted on the STM32MP1 SoC.

\subsection{From Delay-Block to TDC Sensor}

The STM32MP1 SoC comes with three programmable delay-blocks IPs (DLYB \cite{Microelectronics2019}) capable of working with different types of external memories (QSPI, SD, MMC). Their settings can be adjusted depending on the bus speeds of the external memories used. Their initial purpose is to adjust the phase of the clock signal in order to ensure a reliable exchange of data by tuning the clock delay.\par
\begin{figure}[t]
    \centering
    \includegraphics[width=\columnwidth]{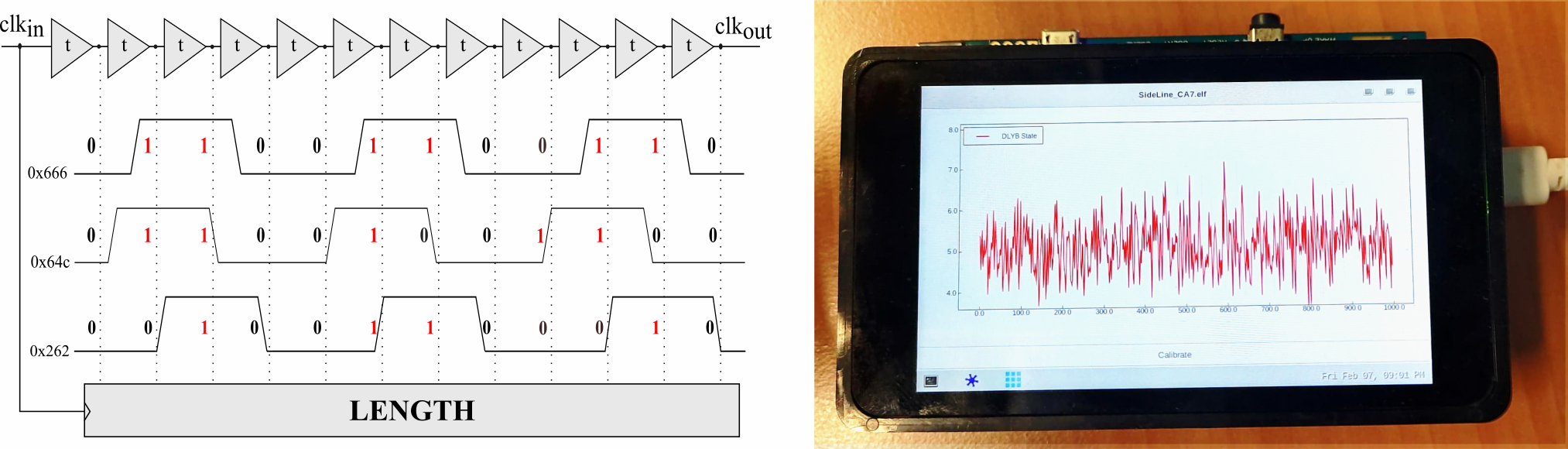}
    \caption{Effect of on-chip voltage variations on the sampled delay values.}
    \label{TDC}
    \vspace{-5mm}
\end{figure}

The left part of figure \ref{TDC} depicts the 12 elements delay-line provided by the STM32MP1 delay-block and the capture register designed to monitor the state of the output nodes of every delay element. 
When a $clk_{in}$ rising edge occurs, the capture register takes a snapshot of the delay-line. This snapshot contains an image (represented as a waveform in Figure \ref{TDC}) of the clock propagation through the delay-line. The propagation delay \textit{t} of the elementary delay elements can be set using a dedicated register. If this delay is set to its minimum the delay-line width (acquisition window) is small. Thus, only a part of the clock signal can be captured. By gradually increasing $t$, the clock signal observation can be extended, possibly to several periods.\par

We leveraged this $t$ parameter to make the delay-block sensitive to on-chip voltage fluctuations. To that end, we took a significant number of delay-line snapshots for each of the 128 possible $t$ delay values. A vast majority of them gave stable results; which means that the captured image remained stable between successive register readings. For a few however, delay variations arose between subsequent captures. This interesting behavior can be explained by (1) on-chip voltage fluctuations that affect the clock propagation time through the delay elements, and (2) by the fact that several delay values \textit{t} naturally position the clock edges in unstable places within the delay line (i.e. in between two delay elements). The left part of figure \ref{TDC} displays three waveforms (delay-line snapshots) obtained with such a $t$ setting. In this configuration, three clock periods stand in the entire delay line. From top to bottom we have: (1) the steady state register waveform which stands as our reference (it outputs a 0x666 reference value), (2) a slowed down waveform that can be obtained due to a supply voltage decrease (it outputs a 0x64c), and (3) an accelerated waveform that can be obtained due to a supply voltage increase (it outputs a 0x262). In our experiments, the three obtained hexadecimal digits are weighted and added to translate into an image of the voltage supply.\par

On the right part of figure \ref{TDC}, a program displays as an oscilloscope the actual delay-line state on the STM32MP1 touchscreen. This way, the actual power consumption noise impact on the delay-block state can be directly observed. To make it possible, the implemented program automatically calibrates the delay-block by testing various delay parameters. For each delay value, it collects multiple delay-line state samples, computes their variance and adopts the calibration that provided the highest variance. Indeed, a higher variance indicates an important delay instability and thus a stronger relationship with voltage fluctuations.

\subsection{Linux-based OpenSSL AES Attack Setup}\label{sub:AttackSetUpDelayBlock}
Similarly to the attack setup described in subsection \ref{sub:AttackSetUpDLL}, we used the OpenSSL AES implementation to evaluate the threat posed by delay-block-based SCAs. The STM32MP1 embeds both a dual core AP and a MCU that makes it possible to test the MCU-vs-AP and AP-vs-MCU attack scenarios introduced in subsection \ref{sub:ThreatModel}. Depending on the scenario, the attack and victim processes were ran either on the AP core or on the MCU core. Here, we consider the MCU-vs-AP attack to describe our attack setup.\par
We use an adapted version of the Zynq-based attack. On the adversary's side (here the MCU), delay-block calibration and use of Hardware Performance Counters (HPCs) were added to the initial algorithm. HPCs are used to accurately time the successive encryptions and to mitigate the de-synchronisation brought by the Linux OS.
For each acquisition, the number of cycles elapsed during the encryption is compared to a maximal limit $Nb_{cycle}$ set by the adversary above which the entire acquisition is discarded.
Prior to the attack, a preliminary test was conducted in order to identify the optimal value for $Nb_{cycle}$ (assuming that a lower number of clock cycles corresponds to a lower number of interrupts). Hence, by launching thousands of AES encryptions, we were able to find a reference number of clock cycles for almost interrupt-free encryptions.
Then, based on this reference, we set a maximal limit $Nb_{cycle}$ beyond which we decided to discard the acquisitions. 
By doing so, at least half of the total acquisitions were retained and used for the subsequent CPA calculations.\par
Regarding the CPA, we embedded it directly within the STM32MP1. This way, we drastically limited the amount of data exported. Moreover, this allowed us to directly plot the results on screen as illustrated in appendix figure \ref{Appendix:cpadisplay}.

\subsection{Delay-block-based SCA Attacks on STM32MP1 SoC}

\begin{figure}[t]
\centering
\includegraphics[width=\columnwidth]{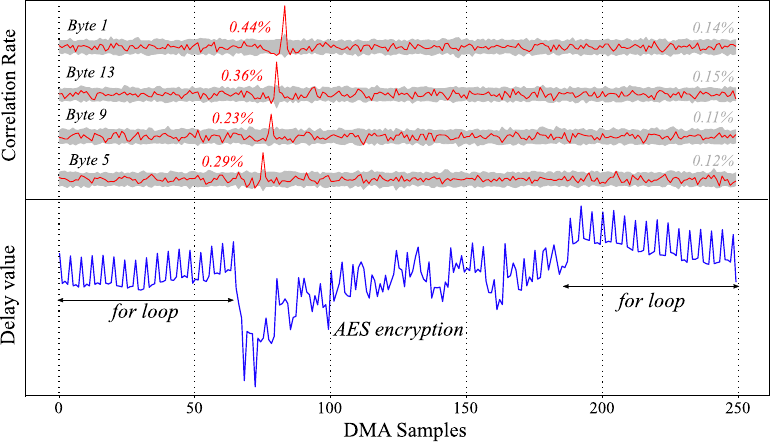}
\caption{AP-vs-MCU attack results: the bottom part represents the averaged AES power consumption, the top part provides the correlation rates as a function of time for four AES key bytes.}\label{CM4_attack}
\vspace{-5mm}
\end{figure}

\textbf{In the AP-vs-MCU attack scenario}, the OpenSSL AES program runs within the STM32MP1 Cortex-M MCU. Using compiler optimization set to -O0, 1,460 clock cycles are required to perform a single AES encryption, that is 7.3\,$\mu$s at the MCU operating frequency (200 MHz).
Figure \ref{CM4_attack} displays in its bottom part the averaged delay values obtained for a time window of 250 DMA samples (or 16.4 $\mu$s) over 10 million acquisitions. The AES encryption, which approximately covers 110 DMA samples, is surrounded by two empty \texttt{for} loops added for visualisation ease. 
The top part of Figure \ref{CM4_attack} provides the CPA correlation rates of four key bytes (of index \#1, \#13, \#9, and \#5) as a function of time. The correct key hypotheses are depicted in red and emerge from the incorrect hypotheses (in grey) between samples 70 and 80. We chose to represent these key bytes because they 
are equally distant regarding the OpenSSL byte computation order: \textit{0 \textbf{5} 10 15 - 4 \textbf{9} 14 3 - 8 \textbf{13} 2 7 - 12 \textbf{1} 6 11}. This explains the regular temporal offset observed between them. 
Based on 10 million encryptions, we achieved a full AES key recovery. 6 bytes were retrieved in the range 0-2M traces, 4 between 2-6M and 6 between 6-10M. The progressive correlation of the eight last AES key bytes (\#8 to \#15) are depicted in Figure \ref{Appendix:AP-vs-MCU} in the appendix.\\

\par
\begin{figure}[t]
\centering
\includegraphics[width=\columnwidth]{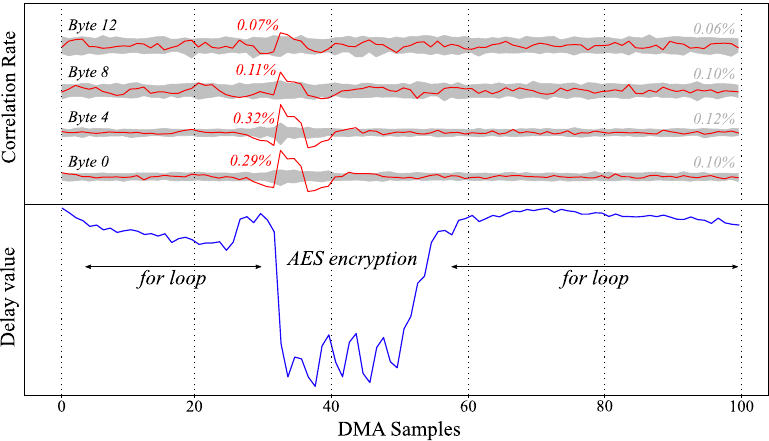}
\caption{MCU-vs-AP attack results: the bottom part represents the averaged AES power consumption. The top part provides the correlation over the time results over four AES key bytes.}
\label{CA7_attack}
\vspace{-5mm}
\end{figure}

\noindent\textbf{In the MCU-vs-AP attack scenario}, the OpenSSL AES program runs in the STM32MP1 Cortex-A7 AP. Using compiler optimization set to -O2, 865 clock cycles are required to perform a single AES encryption, that is 1.33\,$\mu$s at the AP operating frequency (650 MHz). Figure \ref{CA7_attack} displays in its bottom part the averaged delay value obtained for a time window of 100 DMA samples (or 6,6 $\mu$s) over 40 million acquisitions. 
The AES encryption, which approximately covers 20 DMA samples, is surrounded by two empty \texttt{for} loops added for visualisation ease. 
The top part of Figure \ref{CA7_attack} provides the temporal correlation rate of four key bytes as a function of time . The correct key hypotheses are depicted in red and emerge from the incorrect hypotheses (in grey) between samples 30 and 40.
Again, we chose to represent these specific key bytes because they are equally distant in the OpenSSL byte computation order. However, the AES encryption in the AP is faster than that of the MCU (1.33 $\mu$s vs. 7.3 $\mu$s) and the DMA sampling frequency that remained fixed between the two experiments is no longer sufficient to let the temporal offsets appear. This limited sampling frequency partly explains the higher number of acquisitions required to retrieve some key bytes.
For instance, byte \#12 in Figure \ref{CA7_attack}, seems to suffer from the under sampling and gave poorer correlation results (0,07\%) than byte \#4 (0,32\%) or byte \#0 (0,29\%). 
We were able to confirm this assumption through a second experiment where the AES encryption temporal window had been slightly shifted regarding the DMA: the AES leakage was thus sampled at different timings. This experiment gave better results on several key bytes that struggled to emerge in the previous attack.
Based on 40 million encryptions, we achieved a full AES key recovery. 3 bytes were retrieved in the range 0-10M traces, 6 between 10-20M, 2 between 40-30M,  4 between 30-40M. The 13th key byte never completely emerged from the incorrect candidates, but we assume that a simple brute force can be conducted to retrieve its value. The progressive correlation of the first key bytes (0 to 7) are depicted in Figure \ref{Appendix:MCU-vs-AP} in the appendix.

\section{Discussion}\label{sec:Discussion}

Two delay-line-based power measurement techniques, using a DLL or a delay-block were introduced and studied in this research work. 
Because such delay-line-based components are embedded in almost every high-end digital SoC that uses external memories, the threat model we introduced is serious and shall be considered feasible for a large number of complex SoCs. In this section, we discuss performance, additional attack scenarios and potential countermeasures regarding the \textit{SideLine} attack.

\subsection{Performance and Limitations of \textit{SideLine}}


Table \ref{resulttable} summarizes the results obtained for the three attack scenarios considered in this paper. First, an AP-vs-AP attack was performed on a Zynq SoC using DLL-based sensors. As DLLs provide a limited resolution, a large amount of acquisitions were required to integrate enough information for the CPA to succeed (20 million traces required for full AES key recovery). It took around 12 hours to extract the traces, apply post-processing (filtering) and conduct the CPA attack. The lack of resolution also made post-synchronization nearly impossible and thus implied the collection of leakage traces with a constant synchronization. Apart from performances, the DLL was by far the simplest sensor to implement in our experiments, as it only required the reading of a memory-mapped register. However, care must be taken as in certain cases, DLLs may require additional calibration. For instance, some DLLs can either perform delay calibration continuously or at a set of intervals \cite{ARM2003}. Such parameters should be taken into account by the attacker and calibrated if needed. \par  
The second attack proposed in this paper required a preliminary work to properly turn the delay-block into a custom TDC.  
Then, two delay-block-based power SCAs were conducted on a STM32MP1 SoC. The AP-vs-MCU AES attack took around 10 million traces for a full key recovery (trace acquisition and CPA took approximately 9 hours) while the MCU-vs-AP AES attack required 40 million traces (24 hours). We can compare these results to the attack reported in \cite{Gravellier2019} against an OpenSSL AES implementation in an FPGA-based heterogeneous SoC. In this work, FPGA-based TDCs were able to perform a similar attack using only 90,000 traces (FPGA-to-CPU attack). 
FPGAs indeed offer the possibility to design high resolution and high sampling rate sensors which explain the higher efficiency of their attack.
Such a flexibility is obviously not available in ASICs. For instance, even using DMA in our experiments, the maximum sampling rate achieved (16 MHz) was still way under the FPGA-based TDC sampling rate given in \cite{Gravellier2019} (200 MHz).
Additionally delay-blocks also suffer from a poor resolution as evidenced in Figure \ref{Appendix:resolution} in the appendix. 
Despite these limitations, we demonstrated that such an attack is still feasible without using FPGAs and within a reasonable time and number of traces. \par The presence of DLLs and programmable delay-blocks is already mandatory in high-end SoC devices and should become even more prevalent in the future with the constant increase of memory bus speeds. At the same time, their voltage sensing capability will be progressively enhanced as they will need to meet higher performances requirements. This should make \textit{SideLine} even easier to conduct and detrimental for hardware security in the future.\par 


\begin{table*}[t]
\begin{center}
\begingroup
\setlength{\tabcolsep}{8pt} 
\renewcommand{\arraystretch}{1.1} 
\begin{tabular}{cccccc}
\hline
\textbf{Scenario} & Sensor & \textbf{$Nb_{Acq}$}  & \textbf{$freq_{DMA}$} & \textbf{$freq_{Target}$} & Duration\\
\hline
Zynq  AP-vs-AP & DLL & $20$M & $16$ MHz   & $667$ MHz & $\sim12$ h\\
STM32 AP-vs-MCU & DL  & $10$M & $15.2$ MHz & $200$ MHz & $\sim9$ h\\
STM32 MCU-vs-AP & DL  & $40$M & $15.2$ MHz & $650$ MHz & $\sim24$ h\\
\hline
\end{tabular}
\endgroup
\end{center}
\vspace{-2mm}
\caption{Overall delay-line-based power SCA results.}
\label{resulttable}
\vspace{-8mm}
\end{table*}

\subsection{Hardware \& Software Mitigations}\label{sub:counterm}

This section provides some countermeasure guidelines mitigating \textit{SideLine}:\par\noindent
\textbf{Adding SCA Countermeasures:} A simple way to make the victim process more resilient to power SCAs is the addition of software or hardware SCA countermeasures \cite{Zhou2005,Zhang2016}. As mentioned above, one of the main limitations of \textit{SideLine} comes from the low resolution provided by DLL and delay-blocks. This forces the attacker to acquire a huge number of traces (several million in our case) and makes it nearly impossible to re-synchronize SCA traces. On the victim side, software randomization could be a good candidate to efficiently de-synchronize computations and hence to increase significantly the attack difficulty (e.g. adding random delays in T-Table computations for OpenSSL AES). On the monitoring side (delay-line), a straightforward way to mitigate the attack could rely on the addition of phase and frequency jitter to the clock signal used for sampling the delay-line registers.
\par\noindent\textbf{Preventing Delay-Line Access:} Another countermeasure would act at system level by preventing the access to the delay-line registers by unauthorized software entities. Hence, only the OS for instance would have access to this resource. TrustZone could also be used to place DLLs and Delay-blocks in the secure world and make their use by non-secure world impossible in practice. Locking the access to the DMA module or the hardware performance counters would also represent a significant limitation for the attack setup. 
\par\noindent\textbf{Reducing Delay-Line Sampling Rate:} Preventing delay-line access through privilege rights seems insufficient as a malicious attacker or a compromised OS could overpass it (privileges escalation). A hardware way to mitigate the threat would be to limit the delay-block access to a lower sampling rate (e.g. 10KHz). This could be simply achieved by limiting the access rate to the register that stores delay-line information. This way, even if the power consumption monitoring would remain feasible, it will highly affect the delay sensor performances.  
With such a limited sampling rate it would be probably very challenging for an attacker to conduct SCAs on fast encryption algorithms such as AES.

\par\noindent\textbf{Abandoning Delay-Lines in SoCs:} As \textit{SideLine} revealed their potential misuse as power consumption sensors, the delay-line-based components could be removed from SoC devices and instead, be placed directly within the external memory devices. This drastic choice would require the addition of configuration I/Os in external memories to efficiently calibrate the delay-lines but will almost entirely remove the delay-line threat from the SoC die. However, even outside the SoC, the delay-line threat may remain problematic as inter-chip power SCAs have already been shown feasible \cite{Schellenberg2018a}.

\vspace{-2mm}
\section{Conclusion}\label{sec:conclusion}
\vspace{-2mm}
Previous works demonstrated that remote power SCAs were feasible using FPGA-based delay sensors and microcontroller ADC-based sensors. \textit{SideLine} goes further by proving that unsuspected hardware components available in a broad range of high-end SoC devices, can be turned into power consumption measurement units. In this work, we studied two common SoC resources known as delay-locked-loops and delay-blocks and proved their capability to eavesdrop the voltage activity of cryptographic programs running in different processors. Several core-vs-core attack scenarios on application processors and microcontroller units were conducted. For each scenario, we achieved a full key recovery side-channel attack on the publicly available OpenSSL AES implementation. We believe that these findings open a new era for remote power side-channel attacks. \textit{SideLine} has the advantage of being portable on a wide range of devices as it does not requires the presence of specific circuitry (e.g. FPGA). Because \textit{SideLine} feeds upon SoC complexity, we also believe that it represents a major threat for actual high-end SoC security. More importantly this threat is likely to scale up in line with the constant performance improvements in SoCs and memory devices. 





\bibliographystyle{plain}
\bibliography{Algorithme,Remote_FIA,Remote_SCA,Delay_Sensors,Local_SCA,FPGA,Power_Distribution_Network,DLL_DL,TEE,Trojan,Utility,Reference_Manual}

\onecolumn

\section{Appendix}
\vspace{-5mm}

\begin{algorithm}[ht]
\caption{Zynq processor attack, AP\#0 attack pseudo-algorithm}
\begin{algorithmic}

  \STATE \textbf{Input:} $Nb_{acq}$, $Nb_{sample}$
  \STATE DMA$_{init}()$;
  \STATE UART$_{init}()$;
  \WHILE{$Nb_{acq}$ has not been reached}
  \STATE Send AES plaintext to AP\#1;
  \STATE Launch DMA transfer($Nb_{sample}$);
  \STATE Send $Start_{AES}$ to AP\#1;
  \STATE Wait for $End_{AES}$ flag();
  \STATE Wait for $End_{DMA}$ flag();
  \STATE Export samples through UART;
  \ENDWHILE
  
\end{algorithmic}
\label{CA9-0}

\end{algorithm}
\vspace{-1cm}
\begin{algorithm}[ht]
\caption{Zynq processor attack, AP\#1 victim pseudo-algorithm} 
\begin{algorithmic}
  \STATE \textbf{Input:} $AES_{key}, AES_{plaintext}$
  \STATE AES$_{init}()$;
  \WHILE{infinity}
  \STATE Wait for $Start_{AES}$ flag();
  \STATE Get AP\#0 plaintext;
  \STATE OpenSSL AES encrypt();
  \STATE Send $End_{AES}$ flag to AP\#0;
  \STATE Send AES ciphertext to AP\#0;
  \ENDWHILE
\end{algorithmic}
\label{CA9-1}
\end{algorithm}
\vspace{-1cm}

\begin{figure}[ht]
    \centering
    \includegraphics[width=10.5cm]{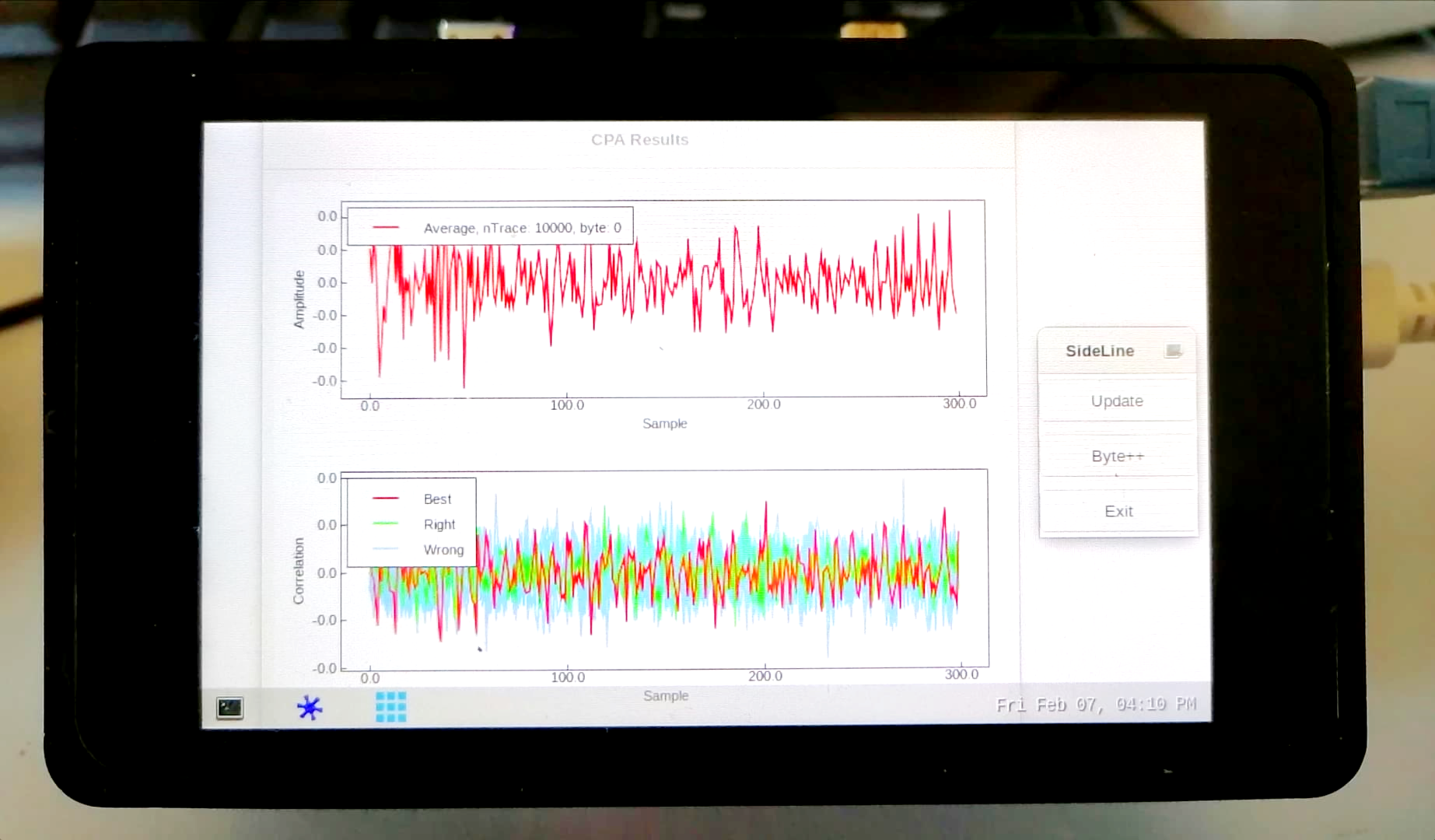}
    \caption{AES traces acquisition, CPA computation and GTK display (implemented for demonstration) are all embedded in the same application running within the STM32MP157-DK2 board.}
    \label{Appendix:cpadisplay}
    \vspace{-8cm}
\end{figure}

\begin{figure}[ht]
    \centering
    \includegraphics[width=10.5cm]{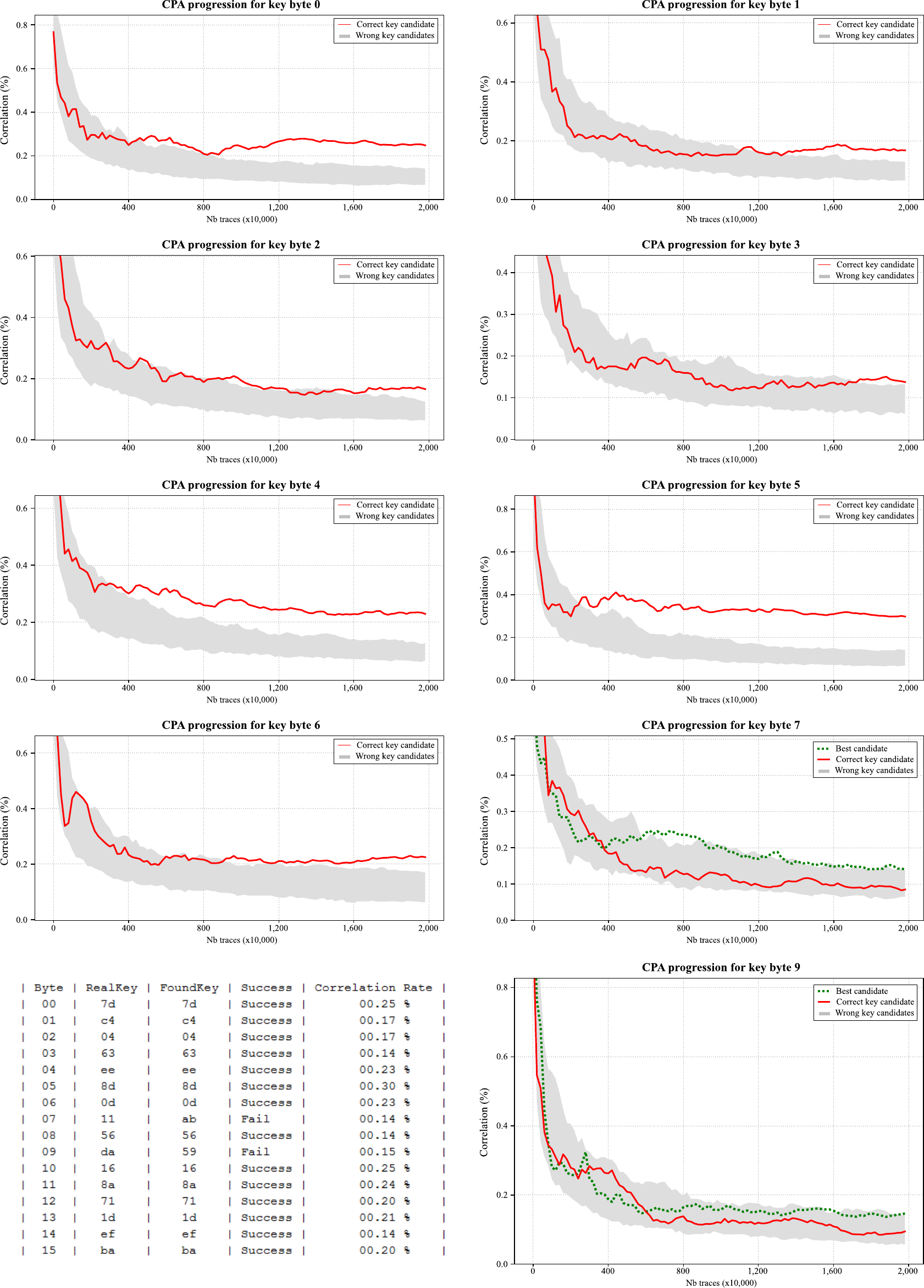}
    \caption{\textbf{ZYNQ AP-vs-AP attack scenario} - The CPA progression (y-axis) over the number of traces (x-axis) is represented for the first 8 AES key bytes. Bytes 7th and 9th which never emerged from the incorrect key candidates are also represented. These CPA results were obtained over 20 million AES encryptions, the correlation rates are provided in the summary table.}
    \label{Appendix:AP-vs-AP}
    \vspace{-3cm}
\end{figure}

\begin{figure*}[ht]
    \centering
    \includegraphics[width=10.5cm]{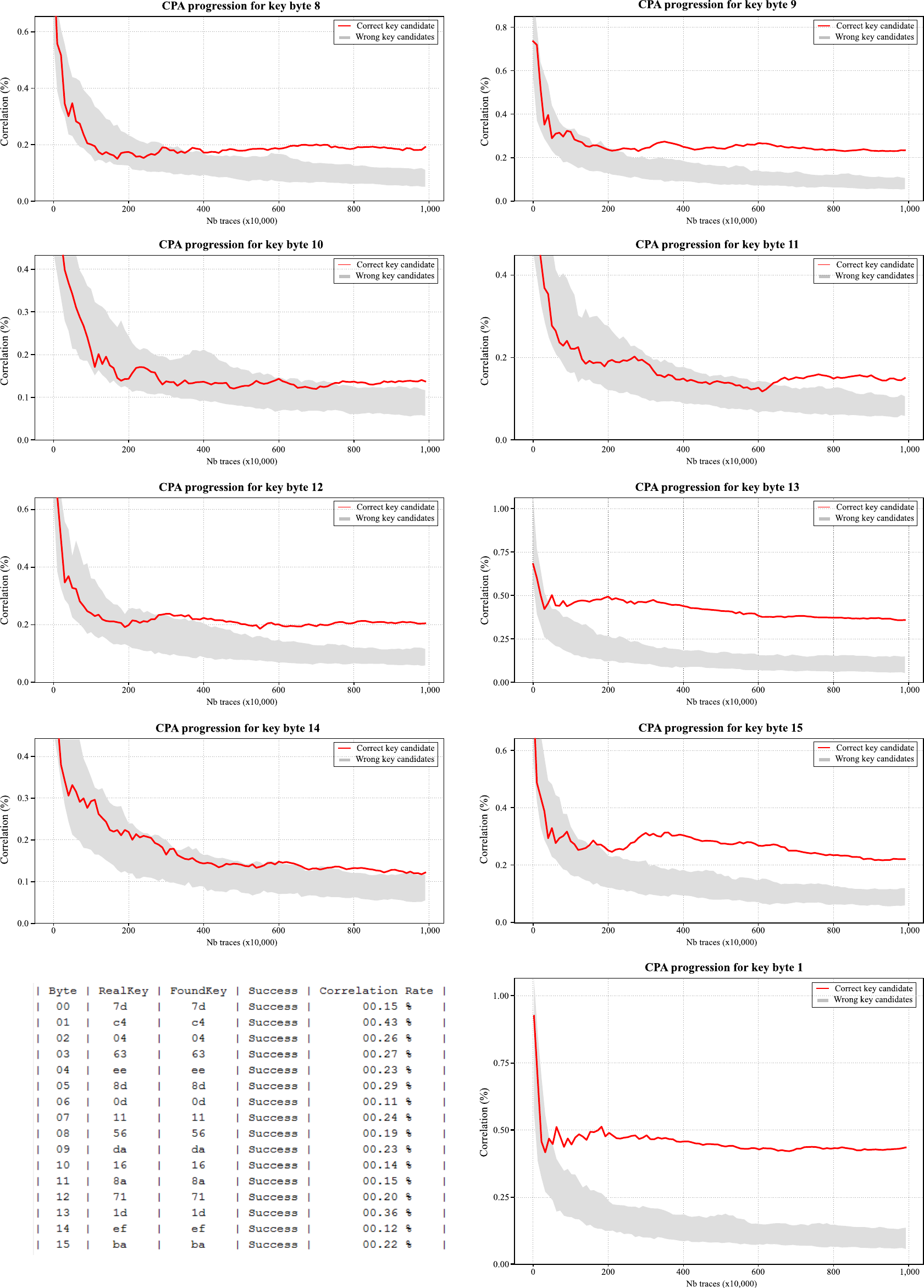}
    \caption{\textbf{STM32MP1 AP-vs-MCU attack scenario} - The CPA progression (y-axis) over the number of traces (x-axis) is represented for the last 8 AES key bytes. The 1st AES key byte is also represented as it provided the best correlation rate. These CPA results were obtained over 10 million AES encryptions, the correlation rates are provided in the summary table.}
    \label{Appendix:AP-vs-MCU}
\end{figure*}

\begin{figure*}[ht]
    \centering
    \includegraphics[width=10.5cm]{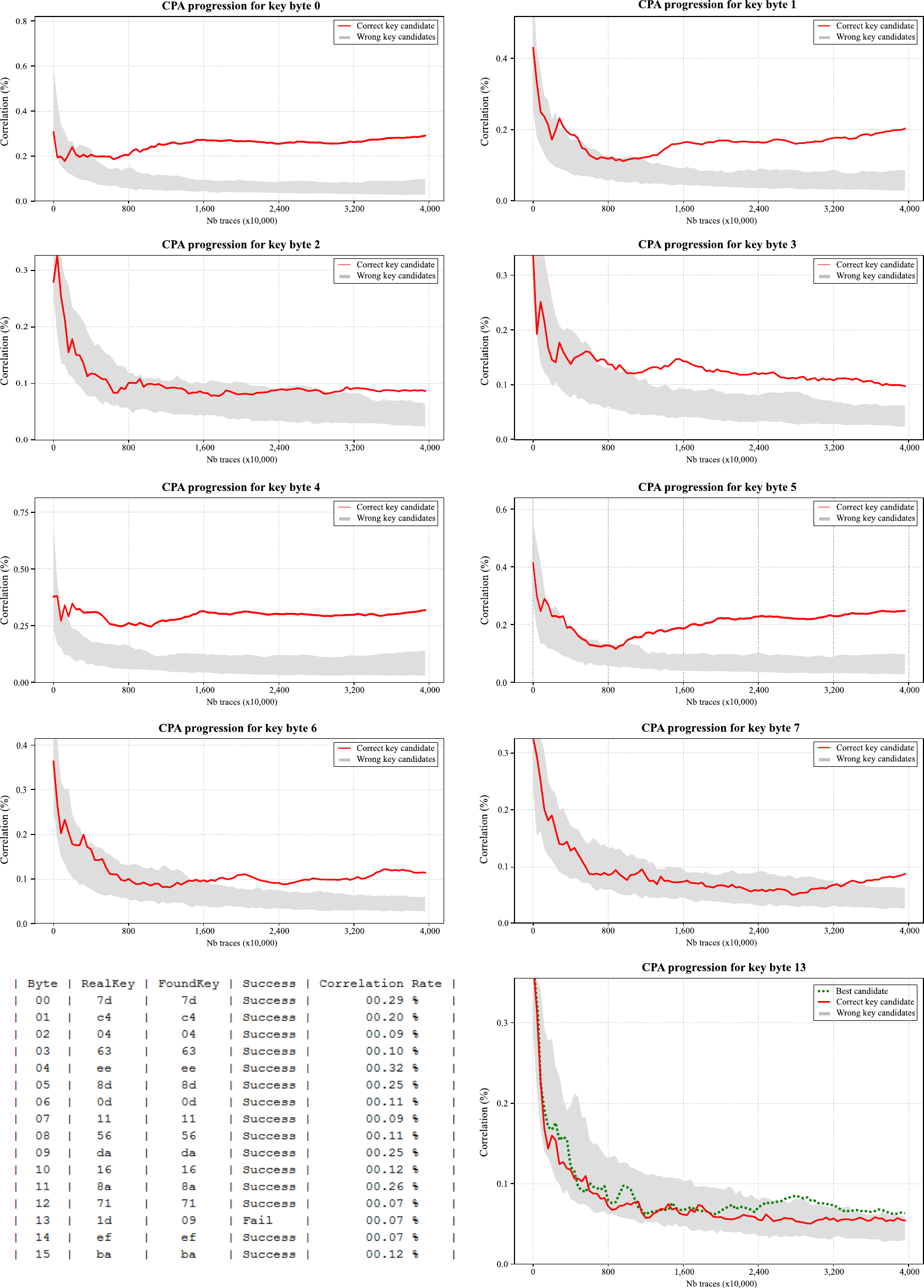}
    \caption{\textbf{STM32MP1 MCU-vs-AP attack scenario} - The CPA progression (y-axis) over the number of traces (x-axis) is represented for the first 8 AES key bytes. Bytes 13th which never emerged from the incorrect key candidates is also represented. These CPA results were obtained over 40 million AES encryptions, the correlation rates are provided in the summary table.}
    \label{Appendix:MCU-vs-AP}
\end{figure*}

\begin{figure*}[ht]
    \centering
    \includegraphics[width=8cm]{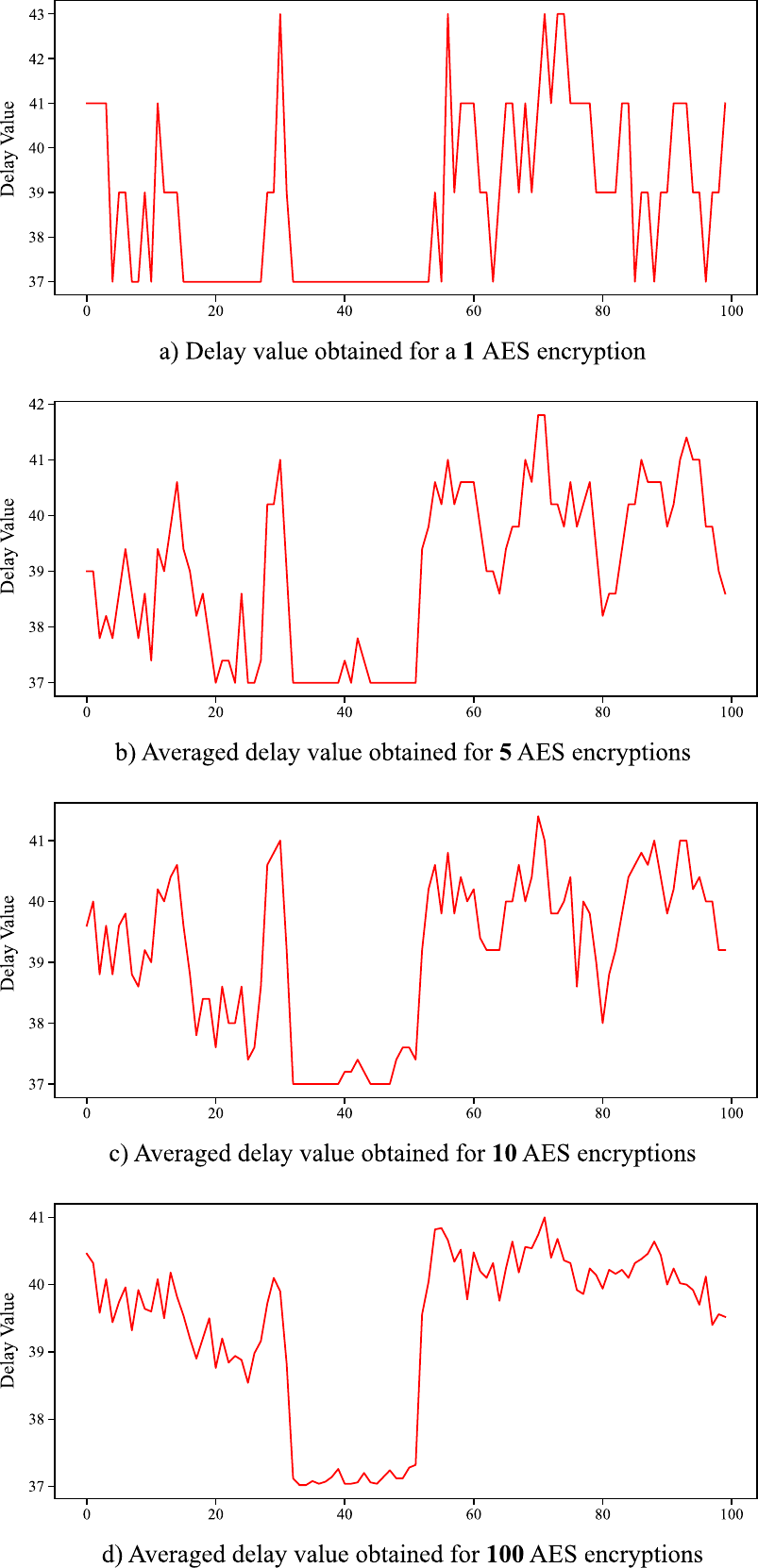}
    \caption{\textbf{STM32MP1 MCU-vs-AP attack} scenario: This figure illustrates the  delay-block resolution limitation when a single AES encryption is acquired (a). This resolution can be virtually increased by averaging a higher number of traces: 5 (b), 10 (c) and 100 (d) traces. }    \label{Appendix:resolution}
\end{figure*}

\end{document}